 \documentclass{aa} 
\usepackage{graphicx}

\newcommand{\kms}{\ifmmode {{\rm km\,s^{-1}}}                 
                  \else {\hbox{{\rm km$\,$s$^{\rm -1}$}}}\fi} 
\def\lo {\ifmmode {\,{\it L}\solar} \else $\,L$\solar\fi}       
\def\my {\ifmmode {\,{\it M}\solar\,{\rm yr^{-1}}}              
        \else {$\,M$\solar$\,$yr$^{\rm -1}$}\fi}

\begin{document}

\title{High resolution millimeter imaging of the proto-planetary
nebula He~3-1475\thanks{Based on observations
carried out with the IRAM Plateau de Bure Interferometer. IRAM is
supported by INSU/CNRS (France), MPG (Germany) and IGN (Spain).}}

    \author{P. J. Huggins\inst{1},
     C. Muthu\inst{2},
     R. Bachiller\inst{2},
     T. Forveille\inst{3, 4},   
     P. Cox\inst{5}
           }

    \offprints{P. J. Huggins}

   \institute{Physics Department, New York University, 4 Washington Place,
              New York, NY 10003, USA
         \and
              IGN Observatorio Astron\'omico Nacional, 
              Apartado 1143, E-28800 Alcal\'a de Henares, Spain
         \and
              Observatoire de Grenoble, B.P. 53X, 38041 Grenoble
              Cedex, France
         \and
              CFHT, PO Box 1597, Kamuela, HI 96743, USA
         \and          
               Institut d'Astrophysique Spatiale, Universit\'e de Paris XI,
               91405 Orsay, France
}

   \date{Received ???; accepted ???}
\authorrunning{Huggins et al. }
\titlerunning{Millimeter imaging of the proto-planetary nebula He~3-1475}

\abstract{
We report high resolution (1\arcsec--2\arcsec) imaging of the CO 2--1
line and the millimeter continuum in the proto-planetary nebula
He~3-1475.  The observations reveal the presence of a massive ($\sim
0.6$~$M_{\sun}$) envelope of molecular gas around the origin of the
remarkable bipolar jet system seen in optical images with the
\emph{HST}. The CO kinematics are well modeled by an expanding,
bi-conical envelope: the prominent, high-velocity ($\sim 50$~\kms)
wings seen in single-dish CO spectra arise where the sides of the
bi-cones are projected along the line of sight. The continuum is
detected at 1.3~mm and 2.6~mm and is due to thermal emission from warm
($\sim 80$~K) circumstellar dust. The structure, kinematics, and
expansion time of the envelope provide strong evidence for entrainment
of the molecular gas by the high velocity jets. The observations
support an evolutionary scenario in which a period of enhanced mass
loss by the central star is followed by the development of the bipolar
jets which burst through the molecular envelope.  The jet-envelope
interactions play a crucial role in shaping the subsequent ionized
nebula.

\keywords{Planetary nebulae: general -- Planetary nebulae: individual:
He~3-1475 -- ISM: jets and outflows -- Stars: AGB and post-AGB} }

\maketitle

\section{Introduction}    

Collimated, bipolar outflows or ``jets'' are an important feature of
the early evolution of planetary nebulae (PNe) that has only recently
been widely recognized.  Many proto-PNe and PNe are now known to show
evidence of bi-polar or multi-polar structures, or point symmetries,
which have been produced by the action of symmetric jets from the
central star (see, e.g., Kastner et al. \cite{ka00}), and these structures
are sufficiently common that possibly most or all PNe pass through this
phase (Sahai \& Trauger \cite{sa98}).

The jets are most active in the early phases of PNe formation, and
their effects can be detected in proto-PNe as high velocity wings in
low angular resolution, molecular line spectra of the neutral
circumstellar envelopes (e.g., Cernicharo et al. \cite{ce89}; Young et
al. \cite{yo92}; Bujarrabal et al. \cite{bu01}). In the few cases
observed at high angular resolution, the wings are seen to arise in
directed outflows of entrained molecular gas (e.g., Cox et
al. \cite{co00}, \cite{co03}).  The origin of the jets that cause the
outflows is not well understood, but the jets clearly have major
effects on the structure and dynamics of the neutral circumstellar
envelopes from which the ionized nebulae form (Huggins et
al. \cite{hu96}), and consequently they play a key role in the early
shaping of PNe. Well studied examples of young PNe that show evidence
for interactions between the jets and the neutral circumstellar gas
include BD+30\degr3639 (Bachiller et al. \cite{ba00}), KjPn~8
(Forveille et al. \cite{fo98}) and M1-16 (Huggins et al. \cite{hu00}).

He~3-1475 (IRAS 17423$-$1755) is an extreme example of a proto-PN with
highly collimated bipolar outflows, first discussed by Riera et
al. (\cite{ri95}) and Bobrowski et al. (\cite{bo95}). It is of special
interest because of the high velocities and remarkable structure seen
in high resolution optical observations made with the HST (Borkowski
et al. \cite{bo97}; Borkowski \& Harrington \cite{bo01};
S\'anchez-Contreras \& Sahai \cite{sa01}; Riera et
al. \cite{ri03}). The outflows terminate in a series of knots that are
point symmetric about the central star, and indicate formation by
episodic jets whose direction is time dependent.


\begin{figure*}[t]
\centering
\resizebox{12cm}{!}{\rotatebox{0}{\includegraphics{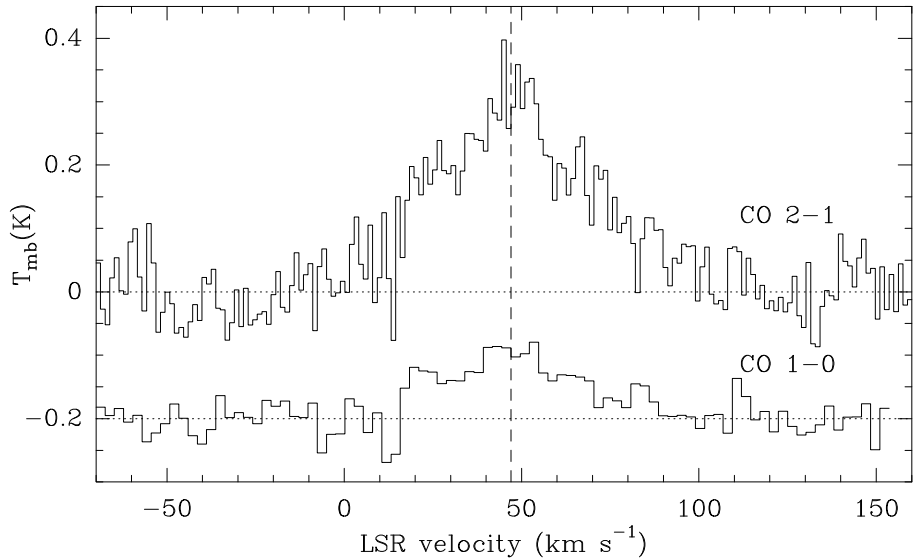}}}
\caption{CO spectra in the 1--0 (115~GHz) and 2--1 (230~GHz) lines
towards He~3-1475, obtained with the IRAM 30~m telescope.
}
\label{fig1}
\end{figure*}


\begin{figure*}
\resizebox{\hsize}{!}{\rotatebox{-90}{\includegraphics{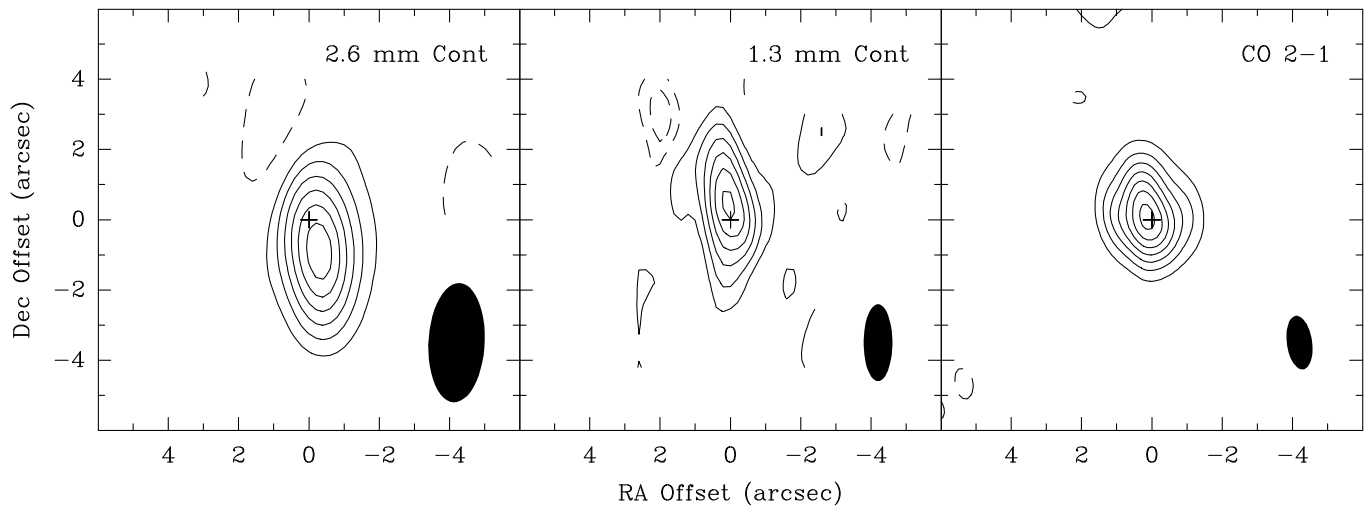}}}
\caption{Millimeter maps of He~3-1475 obtained with the IRAM
interferometer.
\emph{Left:} the continuum at 2.6~mm. \emph{Center:} the continuum at 1.3~mm.
\emph{Right:} the velocity integrated CO 2--1 line.   
The contour intervals are 1~mJy~beam$^{-1}$ (2.6~mm), 2~mJy~beam$^{-1}$ 
(1.3~mm), and 50~K~\kms (CO); dashed contours are negative.
The field center is 17$^{\rm h}$45$^{\rm m}$${14\fs17}$, 
$-$17\degr56\arcmin{47\farcs0} (J2000), and the beam
size is shown in the lower right of each panel.
 }
\label{fig2}
\end{figure*}

Little is known about the neutral circumstellar matter in
He~3-1475. \emph{IRAS} fluxes indicate the presence of a circumstellar
dust envelope (Parthasarathy \& Pottasch \cite{pa89}) which is also seen
in optical images as a dark lane crossing the nebular axis to the
south-east of the central star. OH maser emission at 1667~MHz has been
observed by te Lintel Hekkert (\cite{te91}) and Bobrowski et
al. (\cite{bo95}), and CO spectra have been reported by Knapp et
al. (\cite{kn95}) and Bujarrabal et al. (\cite{bu01}). In this paper
we report high angular resolution observations of the CO emission and
the millimeter continuum to study the relation of the neutral
circumstellar gas to the outflows.


\begin{table*}
\caption[]{CO observations of He~3-1475 made with the IRAM 30~m telescope}
\begin{flushleft}
\begin{tabular}{lllllll}
\noalign{\smallskip}
\hline
\noalign{\smallskip}
   Line    & $V_{\rm o}({\rm LSR})$ & $\Delta V({\rm FWHM})$ & $T_{\rm mb}$  & rms  & $\int
TdV$ & $I$ \\ 
            & (\kms) & (\kms) & (K) & (K) & (K\,\kms) & (K\,\kms\,arcsec$^2$)\\
\noalign{\smallskip}
\hline
\noalign{\smallskip}
1--0 & 47.5 $\pm 1.6$ & 43 $\pm 4$ & 0.10 & 0.02 & 4.7 $\pm 0.4$ & 2580\\
2--1&  47.5 $\pm 0.9$ & 53 $\pm 2$ & 0.28 & 0.04 & 15.9 $\pm0.5$ & 2180\\
\noalign{\smallskip}
\hline
\end{tabular}
\end{flushleft}
\end{table*}

%
\begin{figure}
\resizebox{\hsize}{!}{\rotatebox{-90}{\includegraphics{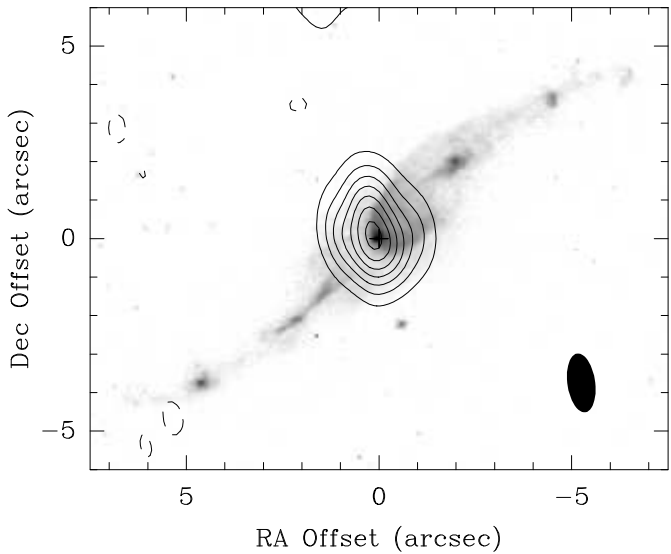}}}
\caption{CO 2--1 map of He~3-1475 superposed on the HST WFPC2 image in
[\ion{N}{ii}] $\lambda$6584 (Borkowski et al. \cite{bo97}). The CO contours
are the same as in Fig.~2. }
\label{fig3}
\end{figure}

\section{Observations}

High sensitivity, single-dish observations of He~3-1475 were made in
the 2.6~mm CO $J=1-0$ (115~GHz) and 1.3~mm $J=2-1$ (230~GHz) lines
using the IRAM 30~m telescope at Pico Veleta, Spain. The observations
were made in September 1997, using $512\times1$~MHz filterbanks. The
half power beam size of the telescope is 11\arcsec\ and 22\arcsec\ at
the frequencies of the 2--1 and 1--0 lines, respectively. The
calibration was made using the chopper wheel technique, and the line
intensities are reported here as main beam brightness temperatures.

High angular resolution observations were made in the CO 1--0 and 2--1 lines
and in the nearby continuum during February and March 1998 using the IRAM
interferometer at Plateau de Bure, France.
The array consisted of five 15~m antennas, equipped with SIS
heterodyne receivers. The observations, centered on He~3-1475, were
made with two configurations of the array, with baselines up to
280~m. The primary beam size of the interferometer is 22\arcsec\ at
1.3~mm and 44\arcsec\ at 2.6~mm. The effective velocity resolution
of the line observations used for the analysis is 8~\kms.  The
continuum observations were made at frequencies of 115.3 and
231.5~GHz, with effective bandwidths of 320 and 640~MHz, respectively.

The RF passband and amplitude were calibrated using 3C273 and the
phase calibration was performed every 20~minutes using J1730$-$130 and
J1830$-$210.  The $uv$ data were Fourier transformed and CLEANed,
using the Clark algorithm and the restored Gaussian clean beam 
is $3\farcs4 \times 1\farcs6$ (${\rm PA} = 177\degr$) at 2.6~mm for
both the line and continuum, and $1\farcs5 \times 0\farcs7$ (${\rm PA} =
8\degr$) and $2\farcs2 \times 0\farcs8$ (${\rm PA} = 0\degr$)
at 1.3~mm for the line and continuum, respectively. The adopted field
center for the maps is 17$^{\rm h}$45$^{\rm m}$${14\fs17}$, 
$-$17\degr56\arcmin{47\farcs0} (J2000).

\section{Results}
The CO 1--0 and 2--1 spectra obtained with the 30~m telescope towards
He~3-1475 are shown in Fig.~1.  A small map was made with 5\arcsec\
spacing around the center position, but the molecular emission was not
found to be extended with respect to the telescope beam.  The
parameters of the lines, based on Gaussian fits to the spectra, and
the corresponding line fluxes are given in Table~1.

The CO 1--0 and 2--1 lines were both detected at high angular
resolution with the interferometer, and the observations recover 97\%
and 72\% of the single-dish fluxes for the 1--0 and 2--1 lines,
respectively. The observed distribution of CO is shown in the velocity
integrated map of the 2--1 emission in the right hand panel of Fig.~2.
The position of the peak CO emission is at $+{0\farcs15}
\pm{0\farcs15}$, $+{0\farcs05} \pm{0\farcs15}$ relative to the map
center, and the emission is extended with respect to the telescope
beam, with a deconvolved source size of $\sim {1\farcs6}$~(FWHM).
The CO distribution is compared with the optical structure of the
nebula in Fig.~3, and the CO kinematics are shown in channel maps and
velocity-position maps, together with model simulations, in Figs.~4--6.
The interferometer observations in the CO 1--0 line produced
essentially the same results as the 2--1 observations, but with a
factor of two lower resolution (because of the longer wavelength), so
are not discussed further.

The millimeter continuum of He~3-1475 was detected at 1.3~mm and
2.6~mm with the interferometer, and maps of the emission are shown in
the center and left hand panels of Fig.~2.  At 2.6~mm the emission is
not extended with respect to the telescope beam, and the measured flux
is $5.3 \pm 1.2$~mJy. At 1.3~mm the observed emission
(${2\farcs8}\pm{0\farcs2} \times {1\farcs1}\pm{0\farcs4}$, ${\rm PA} =
12\degr$) is slightly extended with respect to the beam, and the
measured flux is $31 \pm 4$~mJy.  The position
of the peak emission in the 1.3~mm continuum is at $+{0\farcs04}
\pm{0\farcs05}$, $+{0\farcs40} \pm{0\farcs12}$ relative to the map
center, and is essentially coincident with the position of peak CO
emission, within the uncertainties.
 

\begin{figure*}[t]
\centering
\resizebox{16.5cm}{!}{\rotatebox{-90}{\includegraphics{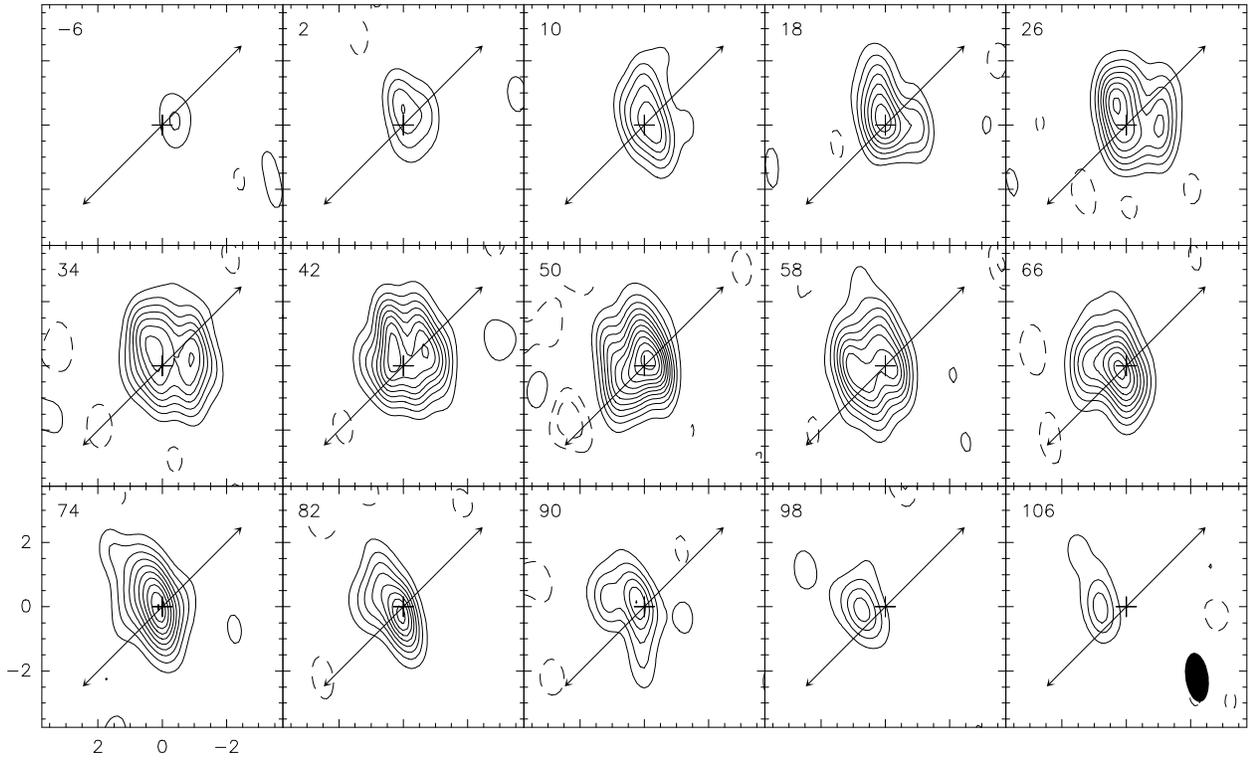}}}
\caption{Channel maps of the CO 2--1 emission observed in He~3-1475.  The
channels are 8~{\kms} wide, and are centered at the velocities given in
the upper left of each panel. The contour interval is 0.5~K; the dashed
contours are negative. The arrows indicate the major (jet) axis 
(${\rm PA} = 135\degr$), and the beam size is shown in the lower right panel.
}
\label{fig4}
\end{figure*}


\begin{figure*}[!]
\centering
\resizebox{16.5cm}{!}{\rotatebox{-90}{\includegraphics{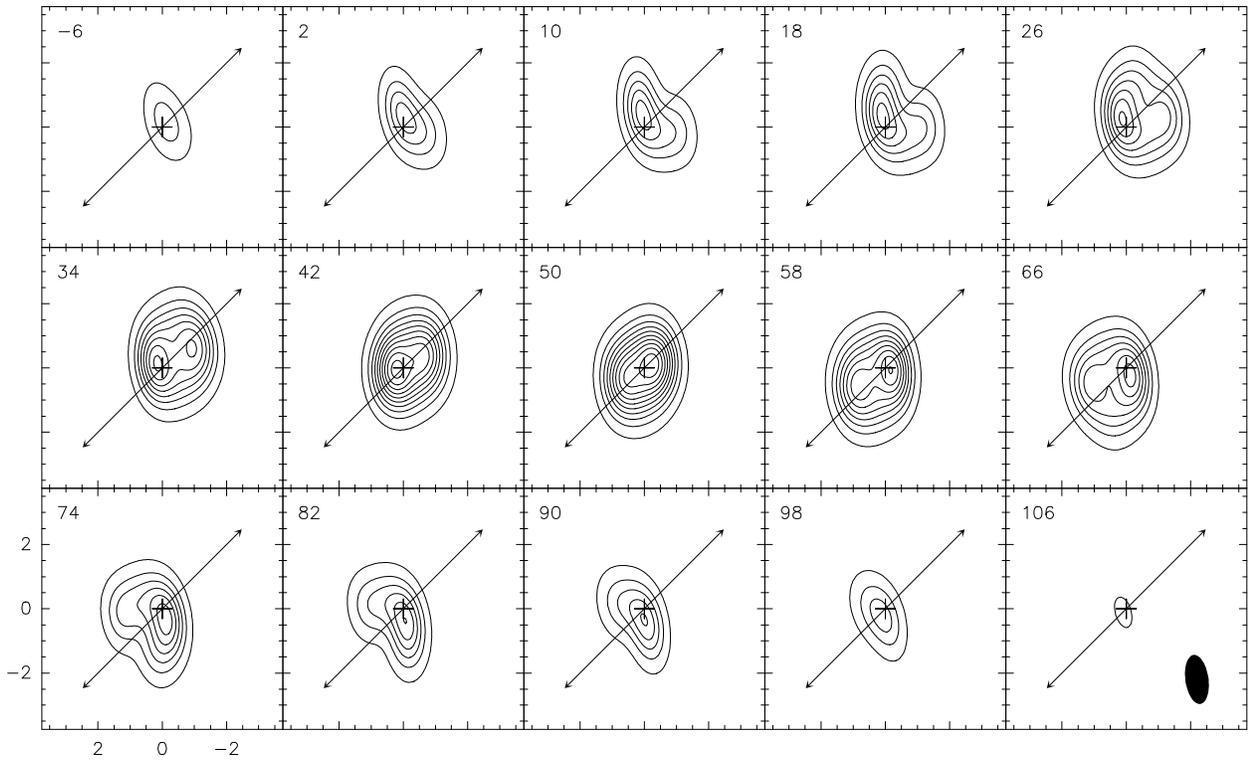}}}
\caption{Theoretical channel maps of the CO 2--1 emission for the
biconical model discussed in the text.  The channels are 8~{\kms}
wide, and are centered at the velocities given in the upper left of
each panel. The contours are from 5\% to 95\% (in steps of 10\%) of
the peak emission.  }
\label{fig5}
\end{figure*}

%
\begin{figure*}
\centering
\resizebox{7.25cm}{!}{\rotatebox{0}{\includegraphics{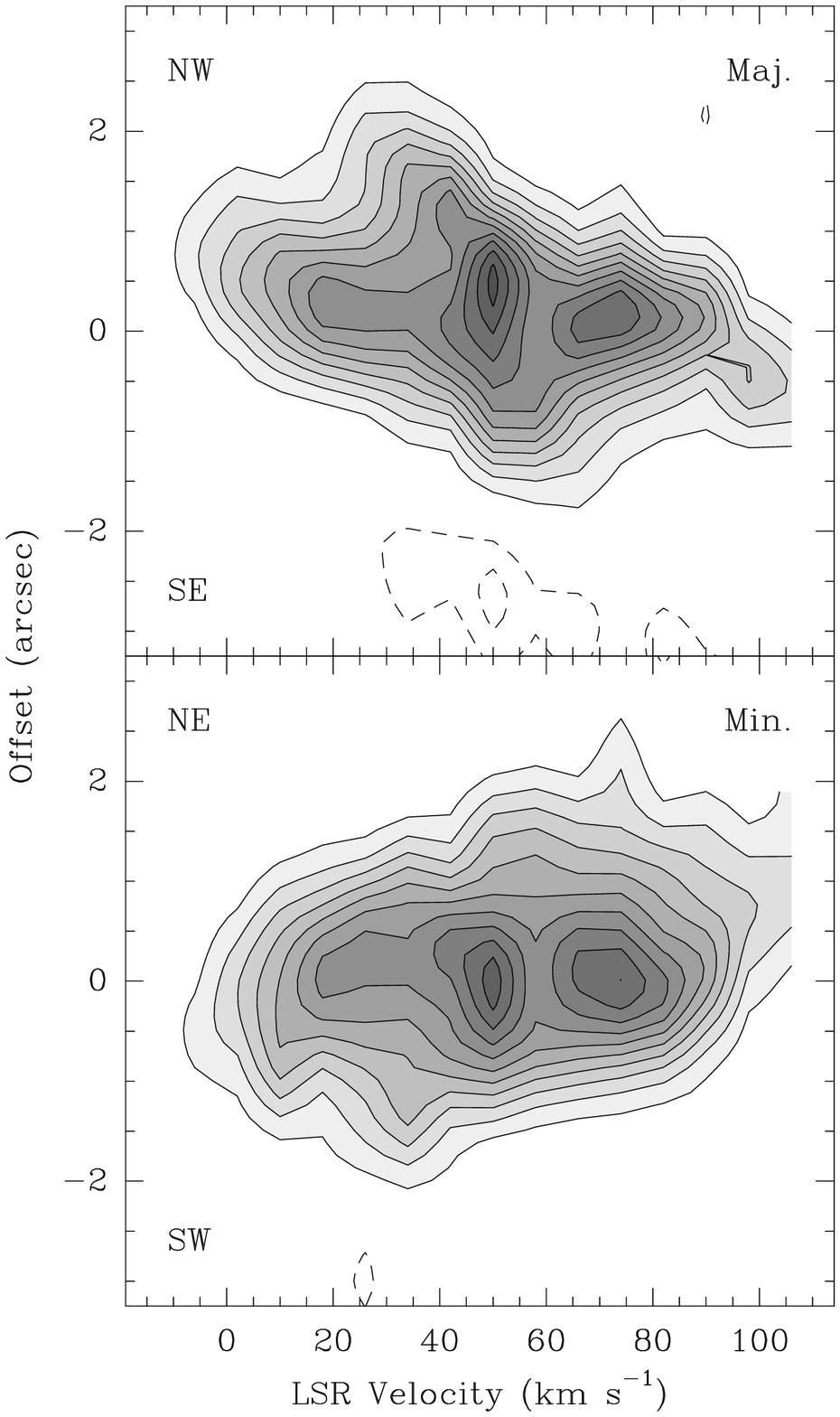}}}
\hspace{1.2cm}
\resizebox{7.25cm}{!}{\rotatebox{0}{\includegraphics{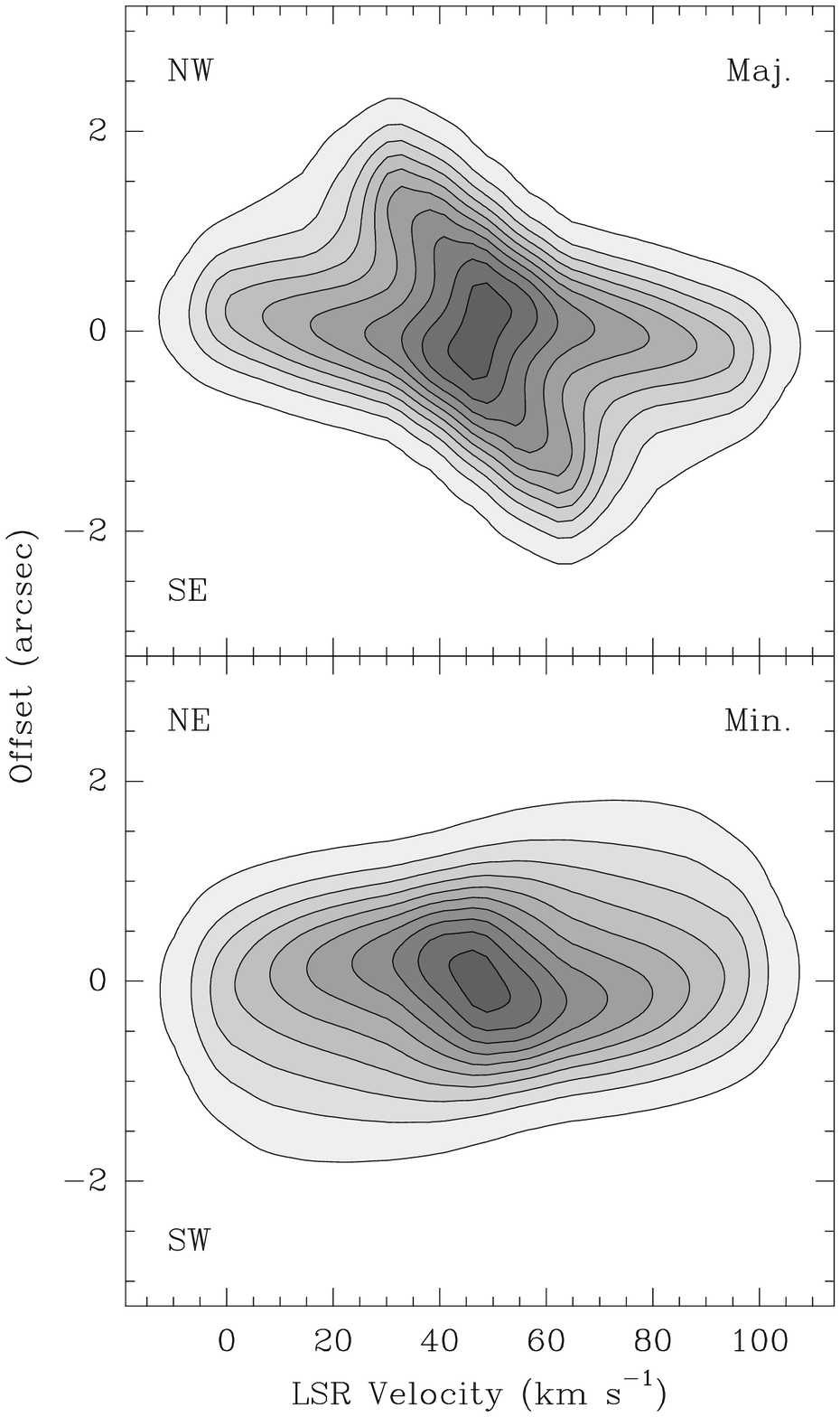}}}
\caption{Position-velocity maps of the CO 2--1 emission in
He~3-1475 along the major ($PA=135\degr$) and minor ($PA=45\degr$) axes.
\emph{Left}: observed data; the contour interval is 0.5~K; dashed
contours are negative. 
\emph{Right}: theoretical maps for the biconical model
discussed in the text. The contours are from 5\% to 95\% (in steps
of 10\%) of the peak emission.}
\label{fig6}
\end{figure*}
\section{Properties of the envelope}

\subsection{Overview}

The CO spectra of He~3-1475 (Fig.~1) are much broader than the
5--15~\kms\ linewidths typically seen in AGB envelopes, and the shapes
of the profiles are different.  They show prominent, high velocity
wings with no clear low velocity component corresponding to an
undisturbed envelope: thus most or all of the molecular gas appears to
participate in the high velocity flow.

From the measurements given in Table~1, the systemic LSR radial
velocity ($V_{\rm o}$) of the molecular gas is well determined to be
$V_{\rm o} = 47.5\pm0.8$~\kms. This corresponds to a heliocentic
velocity of 34.5~\kms, and is in good agreement with the
(heliocentric) velocity of $37.6\pm1.2$~\kms\ recently reported by
Borkowski \& Harrington (\cite{bo01}), based on five stellar lines.

The comparison of the CO map and the HST WFPC2 image in [\ion{N}{ii}]
$\lambda$6584 (Borkowski et al. \cite{bo97}) in Fig.~3 shows that the CO
emission peaks close to the center of the nebula, but is quite limited
in extent compared to the optical image. The molecular emission covers
the central dark lane and extends out along the opening of the bipolar
structure, but does not envelop the extended, collimated jet
system. The characteristic radius of the CO map ({0\farcs8})
corresponds to $7 \times 10^{16}$~cm at a distance of 5.8~kpc, which
we adopt for the distance to He~3-1475 (Riera et al. \cite{ri03}).

\subsection{CO kinematic structure}

Although the CO emission is not very extended with respect to the
telescope beam, the velocity-resolved CO observations provide insights
into the overall kinematic structure of the molecular gas.  The
complete data cube is shown in Fig.~4 in the form of channel maps.
These exhibit an approximate position-velocity symmetry about the
central position and velocity, e.g., the emission in the extreme blue
shifted channels (at $-$6, +2, and +10~\kms) is offset to the NW of
the field center and that of the extreme red channels (+90, +98 and
+106~\kms) is offset to the SE. The maps also exhibit a rough axial
symmetry about the major axis of the optical nebula (the jet axis) at
${\rm PA} = 135\degr$, although the CO intensities are affected by the
shape of telescope beam, which is elliptical and lies at an angle of
53\degr\ with respect to the nebula axis. Thus the arc-like features
in blue channels +10 and +18~\kms\ and red channels +82 and +90~\kms,
which point away from the center, appear stronger where the emission
lies along the primary axis of the beam (see \S 4.3 below).

The velocity-position maps through the data cube along the major and
minor axes shown in Fig.~6, demonstrate that the whole CO structure is
tilted toward us (blue shifted) to the NW and away from us (red
shifted) to the SE along the major axis. This tilt is in the same
sense as that of the optical bipolarity and jets (e.g., Riera et
al. \cite{ri03}).  In addition, the distribution of intensity in the map
along the major axis indicates that the molecular gas forms an
expanding, open ended, bi-conical structure that can also be traced in
the channel maps in Fig.~4.  Thus the extreme, blue shifted emission
(in channels $-$6 and +2~\kms) arises in the lower rim of the cone
facing toward us; this rim lies nearly in the direct line of sight,
and gives rise to the highest blue shifted gas seen in the wings of
the single dish spectra. At intermediate, blue shifted channels
(10--26~\kms), the arc-like structure becomes more prominent as the
channel includes a larger cross section of the cone, and closer to the
systemic velocity (at +42~\kms) the channel also includes the upper
rim of the cone.  At red shifted channels, the overall symmetry is
reversed for the cone facing away from us.

\subsection{Model of the CO kinematics}
In order to quantify our interpretation of the CO data, we have
constructed a simple, biconical model of the CO emission for
comparison with the observations. The model is characterized by the
radius at the equator where the cones intersect ($R_{\rm i}$), the
radius of the open ends ($R_{\rm o}$), the height of each cone ($h$),
and the inclination of the symmetry axis to the line of sight
($i$). We assume a constant thickness ($t$), and a relative density
within the cones that varies as a power law from the center ($\rho
\sim r^{n}$). The emission is taken to be optically thin (see below),
and the expansion is assumed to be homologous, with the radial
velocity $v\sim r$. This velocity law is motivated by the observed
structure in the data cube, and by similar ballistic flows seen in
other proto-PNe (e.g., AFGL~618, Cox et al. \cite{co03}).

From comparison with the observations, the parameters of a best-fit
model are $R_{\rm i}= {0\farcs45}$, $R_{\rm o} = {0\farcs92}$, $h
= {1\farcs5}$, $t={0\farcs5}$, $n = -0.5$, $i=40\degr$, and a velocity
gradient of 31.4~\kms~arcsec$^{-1}$. The linear dimensions are given
in arc seconds on the sky, since they scale with distance.  The
thickness of the walls of the bicones is not actually resolved by the
observations, and is set to a nominal value. The parameter $n$ is
constrained by the relative intensity of the extended CO emission to
that near the center, and the dimensions of the bicones, the velocity
gradient, and the inclination are jointly constrained by the map sizes
at different velocities, and the relative projections of
the sides of the cones in the position-velocity map.  For example, a
general constraint on the inclination and the opening angle of the
bicones is provided by the major axis position-velocity map, where the
blue shifted, lower rim of the cone facing toward us is very close to
the line of sight, and the upper rim is also blue shifted. In this
case, the inclination angle of the symmetry axis $i\la 45\degr$ and
the opening angle of the cone from the center is $\la 45\degr$
(otherwise the upper rim would be red shifted). Similarly the
inclination cannot be much less than $\sim 40\degr$ or the upper rim
would project to higher velocities than are observed.

The results of the best-fit model are shown in Fig.~5 and Fig.~6
(right panel). In spite of the extreme simplicity of the model, it can
account for the main features of the observations. The inclination of
the CO bicones to the line of sight is essentially the same as that of
the optical jets (Borkowski \& Harrington \cite{bo01}), and the biconical
geometry indicates that molecular gas lies around the base of the jet
flows.  As noted above we do not resolve the walls of the bicones, but
the parameter $n$ together with the assumed constant wall thickness
determines the mass distribution in the bicones: for the solution $n =
-0.5$, the mass per unit length projected along the major axis is
approximately constant, which is reasonable for such a flow.

The kinematic model is axi-symmetric but our calculations of the
simulated observations also include the elliptical telescope beam
which is not aligned along this axis. The effects can be seen in
Figs.~5 and 6. They include the asymmetries in intensity of the
arc-like features in the channel maps noted earlier, and a more subtle
effect in a shift from red to blue (from NE to SW) in the minor axis
position-velocity map, which is also apparent in the real data. Two
features of the observations are not reproduced by the model
simulations. The first is a slight channel-to-channel shift in the
position of the emission at high velocities, and this could be
accommodated by some degree of curvature in the sides of the
bicones. The second is slightly higher equatorial emission in the
observations than in the model, which could be accommodated by higher
densities or a modified geometry near the equator.  However, the
emission is probably at least partially thick, especially near the
equator, and we have not included this in the simple model, so fine
tuning with additional parameters would be unwarranted. Nevertheless,
independent of optical depth effects, the basic kinematic structure is
well accounted for by the simple model.

%
\begin{figure}
\resizebox{7.25cm}{!}{\rotatebox{0}{\includegraphics{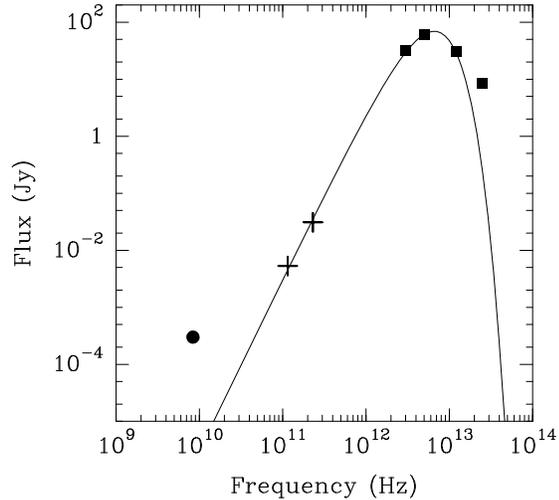}}}
\caption{The infrared--radio spectrum of He3-1475. The flux
measurements are from IRAS (squares), this paper (crosses), and
Knapp et al. (\cite{kn95}) (circle). The smooth curve is for thermal
emission from dust at a temperature of 81~K.
}
\label{fig7}
\end{figure}

\subsection{The circumstellar dust component}

Our millimeter continuum observations fill a large gap in the observed
long wavelength spectrum of He~3-1475, and they provide an estimate of
the mass of circumstellar dust.  As shown in Fig.~7, the only other
long wavelength observations of He~3-1475 are from $IRAS$, and a
single radio observation at 3.6~cm. The centimeter continuum is likely
to be free-free emission from the compact ionized core, and from
comparison with the flux in H\,$\alpha$ Borowski et al. (\cite{bo95}) suggest
that the emission is optically thick.  An optically thick radio
spectrum which varies as $\nu^{2}$ does not, however, extend into the
millimeter region because the 115~GHz (2.6~mm) flux would then be much
higher than we observe. The free-free emission spectrum likely becomes
optically thin and levels off between 10 and 100~GHz.

The millimeter flux increases quite steeply with increasing frequency,
and from the overall shape of the spectrum the main contribution to
the observed 1.3~mm and 2.6~mm flux is probably the long wavelength
tail of the dust emission seen in the infrared.  This is supported by
the fact that we are able to simultaneously provide a good fit to the
25--100~$\mu$m (color corrected) $IRAS$ fluxes and to the observed
millimeter fluxes, with a single temperature dust component.  Formal
fits, varying both the temperature and the index of dust emissivity $p$
(where $Q(\nu) \sim \nu^p$), give $T_{\rm d} = 81\pm4$~K, and $p =
0.99\pm0.07$.
 
With this value for the temperature, we estimate the mass of the cool
dust component using the optically thin expression $M_{\rm d} =
F_{\nu}D^2/(\chi_\nu B_\nu(T))$, where $F_{\nu}$ is the flux,
$\chi_{\nu}$ the emissivity per unit mass, and $B_\nu$ the Planck
function at frequency $\nu$, and $D$ is the distance. Using the $IRAS$
flux at 60~$\mu\/{\rm m}$, $\chi_{60} = 150$~cm$^2$~g$^{-1}$ (Jura
1986), and $D= 5.8$~kpc, we find $M_{\rm d}
=6.4\times10^{-3}$~M$_{\sun}$. A second component with a higher
temperature and a much smaller mass, which we ignore, can account for
the additional, short wavelength emission in the spectrum in Fig.~7.

\subsection{Mass of circumstellar gas}

The mass of circumstellar gas can be estimated from the dust mass
given above, by adopting a value for the gas-to-dust ratio.  For
transition objects with high mass loss rates, the gas-to-dust ratio
appears to be somewhat smaller than for typical AGB stars, and we
adopt the value of $\sim 100$ from Knapp et al. (\cite{kn93}). This yields a
mass of circumstellar gas $M_{\rm g} = 0.64$~M$_{\sun}$.

The CO observations provide a second estimate for the mass of
circumstellar gas. In this case, it is a lower limit because
the low ($\sim 0.9$) CO 2--1/1--0 flux ratio (Table 1) suggests that
the CO lines may be at least partially optically thick, although an
alternative possibility is that the CO is sub-thermally excited.  In
any event we obtain a lower limit to the mass of molecular gas using
the optically thin formula given by Huggins et al. (\cite{hu96}). For a distance
of 5.8~kpc and a representative CO/H$_2$ abundance of $2\times
10^{-4}$ which is commonly assumed for oxygen-rich envelopes (e.g.,
Kahane \& Jura \cite{ka94}), we find $M_{\rm g} \ga 0.19$~M$_{\sun}$,
consistent with the estimate given above.

These values are in accord with a further estimate of the mass of
circumstellar gas, given by Bujarrabal et al. (\cite{bu01}) based on
observations of the $^{13}$CO lines.  This approach minimizes the
effect of line opacity because the lines are likely to be optically
thin, but relies on an assumed value for the $^{13}$CO/H$_2$
abundance. For their assumed value ($2\times10^{-5}$), and our adopted
distance of 5.8~kpc to He~3-1475, $M_{\rm g} = 0.85$~M$_{\sun}$,
consistent with the above values.

\section{Jet-envelope interactions}

Our observations of the molecular emission in He~3-1475 support an
evolutionary scenario in which a period of enhanced mass loss by the
central star is followed by the development of bipolar jets that burst
through the surrounding molecular gas.

\subsection{Mass loss rate of the precursor} 

The current mass of circumstellar molecular gas around He~3-1475 is
substantial (\S 4.5), and it was presumably ejected by the star at a
moderate velocity characteristic of the AGB. For a velocity of $\sim
15$~\kms, the size of the CO map that we observe (\S 3) implies an
ejection time scale of $\la 1,500$~yr, and a corresponding mass loss
rate over this time of $\ga 1.3 \times 10^{-4}$~$M_{\sun}$\,yr$^{-1}$,
using our lower limit on the CO mass.  This mass loss rate is much
larger than typically seen on the AGB, but is characteristic of some
other transition objects (e.g., AFGL~2688, Jura et al. \cite{ju00}), and is
either an intrinsic part of the final evolution of single stars on the
AGB, or the result of binary interactions.

\subsection{Envelope entrainment}

In addition to enhanced mass loss, the kinematics and structure of the
circumstellar gas indicate that most or all of the molecular envelope
has been affected by recent interaction with the jets.  First, the CO
velocities (up to $\sim 50$~\kms) are significantly larger than the
expansion velocities of AGB stars, but are much less than velocities
seen in the ionized gas close to the jet axes (Borkowski \& Harrington
\cite{bo01}; S\'anchez-Contreras \& Sahai \cite{sa01}; Riera et
al. \cite{ri03}).  Second, the observations show that the molecular
gas forms an expanding bi-conical structure around the base of the
optical bipolar flows.  Thus both the structure and kinematics provide
strong evidence for entrainment of the molecular gas.

The kinematic time scales support this view. From proper motion
studies using HST, the expansion time scales ($r/v$ projected on the
plane of the sky) of the ionized knots at $\pm6\arcsec$ from the
center are $\tau_{\rm exp}\sim 450$--550~yr, and for the most distant
knots at $\pm{7\farcs5}$, $\tau_{\rm exp}\sim 600$~yr (Riera et
al. \cite{ri03}; Borkowski \& Harrington \cite{bo01}). For comparison,
the kinematic time scale of the molecular gas is obtained by combining
the (angular) velocity gradient of our kinematic model (\S4.3) and the
adopted distance of 5.8~pc, which give $\tau_{\rm exp}\sim 875$~yr.
Thus the CO outflows and the earliest jets are comparable in age.

There are large velocity gradients from the axes of the jets to the
sides of the CO bicones, as expected if the jets burst through the
envelope.  The maximum CO velocity that we observe is $\sim 50$~\kms,
which is in fact close to the maximum velocity for molecular gas to
survive acceleration in a single shock (Draine et al. \cite{dr83}), although
it is possible for molecules to reform at higher velocities in
post-shocked gas (Hollenbach \& McKee \cite{ho89}; Neufeld \& Dalgarno
\cite{ne89}).  Details of the interaction need further study, but it
seems likely that much of the fast, ionized material seen in the
optical outflows has been produced by the destruction of molecules in
the entrainment process.  Similarly, there may also be an intermediate
component of neutral atomic gas produced by the interactions which
could be detected in species such as \ion{C}{ii}, \ion{C}{i}, and
\ion{O}{i}.

Molecular H$_2$ has been detected in the axial knots themselves
(Harrington et al. \cite{ha00}). This is probably at very high velocity
(although no spectroscopy has yet been done) and the emission is likely
formed in dense, post-shocked gas, possibly under conditions similar
to those in AFGL~618 discussed by Cox et al. (\cite{co03}).

Weak OH maser emission has also been observed toward He~3-1475 (te
Lintel Hekkert \cite{te91}; Bobrowski et al. \cite{bo95}) and is
distributed in multiple maser spots within $\sim {0\farcs8}$ of the
center, with a velocity range $\sim 20$--75~\kms, roughly centered on
the CO systemic velocity.  The kinematic structure of the OH masers in
Fig.~5 of Bobrowski et al. (\cite{bo95}) appears to show a strong
velocity gradient across the envelope which has been remarked on by
others, e.g., Riera et al. (\cite{ri03}).  This apparent gradient is,
however, not real. It is an artifact of ascribing positive and
negative radial offsets to the red and blue components, and the actual
distribution of the maser components shows no clear cut geometry
(Zijlstra et al. \cite{zi01}). Given the OH maser velocities, the
spots are probably located in the inner regions of the CO bicones near
the equatorial plane.

\subsection{Envelope dynamics}

It has long been suspected that stellar radiation pressure may not
power the highest mass loss rates in post-AGB stars (e.g., Knapp et
al. \cite{kn82}), and recent survey work by Bujarrabal et
al. (\cite{bu01}) has quantified this for the molecular outflows seen
as high velocity wings in single dish CO spectra. We follow their
approach in estimating the linear momentum ($P$) and energy ($E$) of
the outflow using the observed values along the line of sight, and
correcting for the inclination. We use the position-velocity map of
Fig.~6 to define the outflow as the emission at velocities $|V-V_{\rm
o}| > 5$~\kms.  For $d=5.8$~kpc and $i=40\degr$ adopted earlier, we
find $P \ga 1\times10^{39}$~g\,cm\,s$^{-1}$ and $E \ga
2\times10^{45}$~erg using the CO emission.  If we use the mass
estimate based on the millimeter continuum and assume a velocity
distribution as in the CO profile, the actual values are $\sim 3$
times larger than these limits. The results are similar to those
obtained by Bujarrabal et al. (\cite{bu01}) for He~3-1475, based on
the $^{13}$CO lines.

The luminosity of He~3-1475 at a distance of 5.8~kpc is $L=
12,600$~$L_{\sun}$ (Riera et al. \cite{ri03}). If radiation pressure from the
star drives the molecular outflows, the time scale to generate the
observed momentum (defined by the equation $\tau (L/c) = P$) is $\ga 2
\times 10^4$~yr.  However, as discussed in \S5.2, the CO expansion
time is only $\sim 875$~yr, and the actual time to accelerate the
molecular gas to the observed velocities is probably less. Thus
radiation pressure is unable to drive the flow by a large margin, as
found by Bujarrabal et al. (\cite{bu01}) for more than 20 cases.  This result
is not unexpected in He~3-1475 because the geometry and kinematics
discussed above provide strong evidence that the molecular gas is
accelerated by entrainment in the jets, and these highly collimated
structures are not likely generated by radiation pressure.

In He~3-1475 the jets appear to have easily penetrated the dense
circumstellar gas close to the star: the molecular gas is peripheral
to the outflows, and is at relatively low velocities compared to the
jet axes.  In these circumstances, it might be expected that the
energy and momentum of the flow estimated from the molecular gas alone
would be a relatively small part of the total, i.e., that the values
given above are strong lower limits.  It is therefore surprising that
the energy and momentum in the fast outflows estimated by Riera et
al. (\cite{ri03}) from optical observations are $P \sim
10^{37}$~g\,cm\,s$^{-1}$ and $E \sim 10^{45}$~erg, which are
\emph{less} than the values we estimate from the molecular emission,
the value of $P$ less by a factor of 100.

This discrepancy requires explanation and we offer two solutions. One
is that the material close to the jet axes is largely neutral. Riera
et al. (\cite{ri03}) have estimated the flow parameters based on a
density determined from [\ion{S}{ii}] line ratios. If the material is
largely neutral, as expected from calculations of jets in YSOs (e.g.,
Safier \cite{sa93}; Shang et al. \cite{sh02}), then the actual
densities could be much larger, and the dynamical quantities
correspondingly higher. A second possibility is that the primordial
jets are in fact components of a wide angle wind. If so, the wide
angle wind might couple well to the molecular gas, with only the
central, highest velocity material breaking through the envelope to
form the visible jets.  These scenarios need to be examined in other
proto-PNe.

\section{Conclusions}

The millimeter observations reported here provide basic information on
the properties of the neutral circumstellar matter around the
remarkable proto-planetary nebula He~3-1475. The observations also
underscore the importance of neutral circumstellar gas in the early
development of PNe.

He~3-1475 is surrounded by a massive circumstellar envelope ($\ga
0.2$~$M_{\sun}$ and $\sim 0.6$~$M_{\sun}$ from our CO and continuum
observations, respectively), which has only recently been ejected by
the central star at a high mass loss rate ($\ga 1
\times10^{-4}$~$M_{\sun}$\,yr$^{-1}$). The structure and kinematics of
the CO emission are well modeled with an expanding bi-conical
envelope, and lead to the conclusion that the molecular gas has been
entrained in the sides of the jets as they burst through the
envelope. The expansion time scales of the CO emission and the jets
support this view.

Although He~3-1475 is an extreme object on account of the very high
velocity of the jets and its well developed point symmetric structure,
the evolutionary scenario outlined above is very similar to other
newly forming PNe we have studied at high resolution.  AFGL~2688 (Cox
et al. \cite{co00}), CRL~618 (Cox et al. \cite{co03}); M1-16 (Huggins
et al. \cite{hu00}), KjPn~8 (Forveille et al. \cite{fo98}), and
NGC~7027 (Cox et al. \cite{co02}) form an approximate evolutionary
sequence in which the ionized nebula turns on and becomes more
dominant, and in each case there are prominent multiple jets, or
single jets which have changed direction.  He~3-1475 clearly belongs
with this class.  Besides constraining the origin of the jets and the
physics of jet-envelope interactions, the observations of these
objects demonstrate the importance of jets in the early shaping of the
neutral circumstellar envelopes which play a key role in determining
the morphology of the mature PNe.

\begin{acknowledgements}
We thank Drs.~M. Bobrowski, J.~P. Harrington, and A. Zijlstra for
useful discussions, and J.~P. Harrington for the image used in Fig.~3.
We also thank the IRAM staff at Plateau de Bure for making the
observations, and the staff at Grenoble for help with the data. This
work was supported in part by NSF grants AST 99-86159 and AST 03-07277
(to P.J.H.) and the Spanish DGES grant AYA2000-927 (to R.B.).
      
\end{acknowledgements}


\begin{thebibliography}{}

\bibitem[2000]{ba00}
Bachiller, R., Forveille, T., Huggins, P. J., Cox, P., \&
Maillard, J. -P. 2000, A\&A, 353, L5
\bibitem[1995]{bo95}
Bobrowski, M., Zijlstra, A. A., Grebel, E. K., et al. 1995, ApJ, 446, L89
\bibitem[1997]{bo97}
Borkowski, K. J., Blondin, J. M., \& Harrington, J. P. 1997, ApJ, 482,
L97
\bibitem[2001]{bo01}
Borkowski, K. J., \& Harrington, J. P. 2001, ApJ, 550, 778
\bibitem[2001]{bu01}
Bujarrabal, V., Castro-Carrizo, A., Alcolea, J., \& S\'anchez Contreras,
C. 2001, A\&A, 377, 868
\bibitem[1989]{ce89} Cernicharo, J., Guelin, M., Penalver, J.,
Martin-Pintado, J., \& Mauersberger, R.  1989, A\&A, 222, L1
\bibitem[2000]{co00}
Cox, P., Lucas, R., Huggins, P. J., et al.  2000, A\&A, 353, L25
\bibitem[2002]{co02}
Cox, P., Huggins, P. J., Maillard, J. -P., et al. 2002, A\&A, 384, 603
\bibitem[2003]{co03}
Cox, P., Huggins, P. J., Maillard, J. -P., et al. 2003, ApJ, 586, L87
\bibitem[1983]{dr83}
Draine, B. T., Roberge, W. G., \& Dalgarno, A. 1983, ApJ, 264, 485
\bibitem[1998]{fo98}
Forveille, T., Huggins, P. J., Bachiller, R., \& Cox, P.  1998, ApJ, 495, L111
\bibitem[2000]{ha00}
Harrington, J.  P., \&  Borkowski, K. J. 2000, 
 in Asymmetrical Planetary Nebulae II: From Origins to
Microstructures, ed. J. H. Kastner, N. Soker, \& S. Rappaport, ASP
Conf. Ser., 199, 383
\bibitem[1989]{ho89}
Hollenbach, D., \& McKee, C. F. 1989, ApJ, 342, 306
\bibitem[1996]{hu96}
Huggins, P. J., Bachiller, R., Cox P., \& Forveille, T. 1996, A\&A, 315, 284 
\bibitem[2000]{hu00}
Huggins, P. J., Forveille, T., Bachiller, R., \& Cox, P.
2000, ApJ, 544, 889
\bibitem[1986]{ju86}
Jura, M. 1986, ApJ, 303, 327
\bibitem[2000]{ju00}
Jura, M., Turner, J. L., Van Dyk, S., \& Knapp, G. R. 2000, ApJ, 528, L105
\bibitem[1994]{ka94}
Kahane, C., \&  Jura, M. 1994, A\&A, 290, 183
\bibitem[2000]{ka00}
Kastner, J. H., Soker, N., \& Rappaport, S. 2000,
Asymmetrical Planetary Nebulae II: From Origins to Microstructures, 
ed. J. H. Kastner, N. Soker, \& S. Rappaport, ASP
Conf. Ser., 199
\bibitem[1982]{kn82}
Knapp, G. R., Phillips, T. G., Leighton, R. B., et al. 1982, ApJ, 252,
616
\bibitem[1993]{kn93}
Knapp, G. R., Sandell, G., \& Robson, E. I. 1993, ApJS, 88, 173
\bibitem[1995]{kn95}
Knapp, G. R., Bowers, P. F., Young, K., \& Phillips, T. G. 1995, ApJ,
455, 293
\bibitem[1989]{ne89}
Neufeld, D. A., \& Dalgarno, A. 1989, ApJ, 340, 869
\bibitem[1989]{pa89}
Parthasarathy, M., \& Pottasch, S. R. 1989, A\&A, 225, 521
\bibitem[1995]{ri95}
Riera, A., Garcia-Lario, P., Manchado, A., Pottasch, S. R., \& Raga,
A. C. 1995, A\&A, 302, 137
\bibitem[2003]{ri03}
Riera, A., Garcia-Lario, P., Manchado, A., Bobrowsky, M., \&
Estalella, R. 2003, A\&A, 401, 1039
\bibitem[1993]{sa93}
Safier, P. 1993, ApJ, 408, 115 
\bibitem[1998]{sa98}
Sahai R., \& Trauger J. T. 1998, AJ, 116, 1357
\bibitem[2001]{sa01}
S\'anchez Contreras, C., \& Sahai, R. 2001, ApJ, 553, L173
\bibitem[2002]{sh02}
Shang, H., Glassgold, A. E., Shu, F. H., \& Lizano, S.
2002, ApJ, 564, 853
\bibitem[1991]{te91}
te Lintel Hekkert, P. 1991, A\&A, 248, 209
\bibitem[1992]{yo92}
Young, K., Serabyn, G., Phillips, T. G., et al. 1992, ApJ, 385, 265
\bibitem[2001]{zi01}
Zijlstra, A. A., Chapman, J. M., te Lintel Hekkert, P., et al. 2001,
MNRAS, 322, 280 
\end{thebibliography}
\end{document}